# Ad Hoc Protocols Via Multi Agent Based Tools

Ali Bazghandi[1], Mehdi Bazghandi[2]
[1] School of Computer Engineering, Shahrood University of Technology
Shhahrood, Iran

[2] Islamic Azad University, Mashhad Branch
Mashhad, Iran

**Abstract**

*The purpose of this paper is investigating behaviors of Ad Hoc protocols in Agent-based simulation environments. First we bring brief introduction about agents and Ad Hoc networks. We introduce some agent-based simulation tools like NS-2. Then we focus on two protocols, which are Ad Hoc On-demand Multipath Distance Vector (AODV) and Destination Sequenced Distance Vector (DSDV). At the end, we bring simulation results and discuss about their reasons*

## 1. Introduction

A) What is an agent?

An *agent* is an animate entity that is capable of doing something on purpose. That definition is broad enough to include humans and other animals, the subjects of verbs that express actions, and the computerized robots and soft bots. But it depends on other words whose meanings are just as problematical: *animate, capable, doing,* and *purpose*. The task of defining those words raises questions that involve almost every other aspect of ontology.

- *Animate*. Literally, an animate entity is one that has an *anima* or soul. But *anima* is the Latin translation of Aristotle's word *psychê*, which had a much broader meaning than the English word *soul*. Aristotle defined a hierarchy ranging from a vegetative psyche for plants to a rational psyche for humans. The first question is whether

  Aristotle's hierarchy of psyches can accommodate the modern robots and soft bots.

- *Capable*. The agent of a verb plays that role only as long as the action persists, but an entity can also be considered an agent if it has the power to perform some action whether or not it actually does. Formalizing that notion of power raises questions about modality, potentiality, dispositions, and counterfactuals that have been discussed in philosophy for centuries.

- *Doing*. The verb *do* sounds as simple as two other little verbs *be* and *have*. But like those verbs, its dictionary entry has one of the largest number of senses of any word in the English language. A common feature of all those senses is causality and purpose: some agent for some purpose causes some process to occur. This feature not only creates a cyclic dependency of *doing* on *agent*, it also introduces the notions of causality, process, and occurrence.

- *Purpose*. In the <u>top-level ontology</u>, purpose is defined as an intention of some agent that determines the interaction of entities in a situation. That is consistent with the definition of an agent as an entity that does something on purpose, but the circularity makes it impossible to give a closed-form definition of either term.

B) Ad Hoc networks

The term "Ad Hoc" implies that this network is a network established for a special, often extemporaneous service customized to applications. So, the typical Ad Hoc network is set up for a limited period of time. The protocols are tuned to the particular application (e.g., send a video stream across the battlefield; find out if a fire has started in the forest; establish a videoconference among 3 teams engaged in a rescue effort). The application may be mobile and the environment may change dynamically. Consequently, the Ad Hoc protocols must self-configure to adjust to environment, traffic and mission changes.

What emerges from these characteristics if the vision of an extremely flexible, malleable and yet robust and formidable network architecture. An architecture that can be used to monitor the habits of birds in their natural habitat, and which, in other circumstances, can be structured to launch deadly attacks onto unsuspecting enemies.

The complexity of mobile Ad Hoc network designs has challenged generations of researchers since the 70's, Thanks in part to the advances in radio technology, major success have been reported in military as well as civilian applications on this front (eg, battlefield, disaster recovery, homeland defense, etc). At first look, these applications are mutually exclusive with the notion of "infrastructure networks and the Internet" on which most **commercial applications** rely. This is in part the reason why the Ad Hoc network technology has had a hard time transitioning to commercial scenarios and touching people's everyday lives.

This may soon change, however. An emerging concept that will reverse this trend is the notion of "**opportunistic Ad Hoc networking**". An opportunistic Ad Hoc subnet connects to the Internet via "wireless infrastructure" links like 802.11 or 2.5/3G, extending the reach and flexibility of such links. This could be beneficial, for example, in indoor environments to interconnect out of reach devices; in urban environments to establish public wireless meshes which include not only fixed access point but also vehicles and pedestrians, and; in Campus environments to interconnect groups of roaming students and researchers via the Internet. It appear thus that after more than 30 years of independent evolution, Ad Hoc networking will get a new spin and wired Internet and Ad Hoc networks will finally come together to produce viable commercial applications.

## 2. Ad-Hoc protocols

### 2.1 Routing in Ad Hoc networks

Developing support for routing is one of the most significant challenges in Ad Hoc networks and is critical for the basic network operations. Certain unique combinations of characteristics make routing in Ad Hoc networks interesting. First, nodes in an Ad Hoc network are allowed to move in an uncontrolled manner. Such node mobility results in a highly dynamic network with rapid topological changes causing frequent route failures. A good routing protocol for this network environment has to dynamically adapt to the changing network topology. Second, the underlying wireless channel provides much lower and more variable bandwidth than wired networks. The wireless channel working as a shared medium makes available bandwidth per node even lower. So routing protocols should be bandwidth-efficient by expending a minimal overhead for computing routes so that much of the remaining bandwidth is available for the actual data communication. Third, nodes run on batteries which have limited energy supply.

In order for nodes to stay and communicate for longer periods, it is desirable that a routing protocol be energy-efficient as well. This also provides also another reason why overheads must be kept low. Thus, routing protocols must meet the conflicting goals of dynamic adaptation and low overhead to deliver good overall performance.

Routing protocols developed for wired networks such as the wired Internet are inadequate here as they not only assume mostly fixed topology but also have high overheads. This has lead to several routing proposals specifically targeted for Ad Hoc networks. While some of these proposals are optimized variants of protocols originally designed for wired networks, the rest adopt new paradigms such as on-demand routing, where routes are maintained "reactively" only when needed. This is in contrast with the traditional, proactive Internet-based protocols. Other new paradigms also have emerged – for example, exploiting location information fro routing, and energy-efficient routing.

### 2.2 Types of Ad Hoc protocols

There are three types protocols in Ad Hoc networks:
1. Table-Driven routing protocol:
    a. Proactive!!
    b. Continuously evaluate the routes
    c. Attempt to maintain consistent, up-to-date routing information
        i. When a route is needed, one may be ready immediately
    d. When the network topology changes
        i. The protocol responds by propagating updates throughout the network to maintain a consistent view

    Famous protocols in this type, are DSDV, CGSR and WRP.
2. On-Demand routing protocol:
    a. Reactive!!
    b. On-Demand style: create routes only when it is desired by the source node
        i. Route discovery: invoke a route-determination procedure, the procedure is terminated when
            1. A route has been found
            2. No route is found after all route permutations are examined
        ii. Route maintained by a route maintenance procedure until
            1. Inaccessible along every path from the source
            2. No longer desired

    c. Longer delay: sometimes a route may not be ready for use immediately when data packets come

Famous protocols in this type, are AODV, DSR, TORA, SSR and RDMAR.

3. Hybrid routing protocol:
   a. Hybrid of table-driven and On-Demand!!
   b. From each node, there is a concept of "zone".
      i. Within each zone, the routing is performed in a table-driven manner (proactive), similar to DSDV.
      ii. However, a node does not try to keep global routing information.
   c. For inter-zone routing, on-demand routing is used.
      i. This is similar to DSR.

Famous protocol in this type, is ZRP.

## 2.3) DSDV routing protocol

**Destination-Sequenced Distance-Vector (DSDV)** was one of the earliest protocols developed for Ad Hoc networks. Primarily design goal of DSDV was to develop a protocol that preserves the simplicity of RIP, while guaranteeing loop freedom. It is well known that Distributed Bellman-Ford (DBF), the basic distance vector protocol, suffers from both short-term and long-term routing loops (the *counting-to-infinity* problem) and thus exhibits poor convergence in the presence of link failures. Note that RIP is DBF with the addition of two Ad Hoc techniques (split-horizon and poisoned-reverse) to prevent two hop loops. The variants of DBF proposed to prevent loops (Merlin-Segall, Jaffe-Moss, and DUAL), however, involve complex inter-nodal coordination. Because of inter-nodal coordination, the overheads of these proposals are much higher than basic DBF and match that of link-state protocols using flooding to disseminate link-state updates; so, these protocols are effective only when topology changes are rare.

The main idea in DSDV is the use of destination sequence numbers to achieve loop freedom without any inter-nodal coordination. Every node maintains a monotonically increasing sequence number for itself. It also maintains the highest known sequence number for each destination in the routing table (called "destination sequence numbers"). The distance/metric information for every destination, typically exchanged via routing updates among neighbors in distance-vector protocols, is tagged with the corresponding destination sequence number. These sequence numbers are used to determine the relative freshness of distance information generated by two nodes for the same destination (the node with a higher destination sequence number has the more recent information). Routing loops are prevented by maintaining an invariant that destination sequence numbers along any valid route monotonically increase toward the destination.

DSDV also uses triggered incremental routing updates between periodic full updates to quickly propagate information about route changes. In DSDV, like in DBF, a node may receive a route with a longer hop count earlier than the one with the smallest hop count. Therefore, always propagating distance information immediately upon change can trigger many updates that will ripple through the network, resulting in a huge overhead. So, DSDV estimates route settling time (time it takes to get the route with the shortest distance after getting the route with a higher distance) based on past history and uses it to avoid propagating every improvement in distance information.

## D) AODV routing protocol

**Ad Hoc On-demand Distance Vector (AODV)** shares DSR's on demand characteristics in that it also discovers routes on an *"as needed"* basis via a similar route discovery process. However, AODV adopts a very different mechanism to maintain routing information. It uses traditional routing tables, one entry per destination. This is in contrast to DSR, which can maintain multiple route cache entries for each destination. Without source routing, AODV relies on routing table entries to propagate a RREP back to the source and, subsequently, to route data packets to the destination. AODV uses destination sequence numbers as in DSDV to prevent routing loops and to determine freshness of routing information. These sequence numbers are carried by all routing packets.

The absence of source routing and promiscuous listening allows AODV to gather only a very limited amount of routing information with each route discovery. Besides, AODV is conservative in dealing with stale routes. It uses the sequence numbers to infer the freshness of routing information and nodes maintain only the route information for a destination corresponding to the latest known sequence number; routes with older sequence numbers are discarded even though they may still be valid. AODV also uses a timer-based route expiry mechanism to promptly purge stale routes. Again if a low value is chosen for the timeout, valid routes may be needlessly discarded.

In AODV, each node maintains at most one route per destination and as a result, the destination replies only once to the first arriving request during a route discovery. Being a single path protocol, it has to invoke a new route discovery whenever the only path from the source to the destination fails. When topology changes frequently, route discovery needs to be initiated often which can be very inefficient since route discovery flood is associated with significant latency and overhead. To overcome this limitation, we have proposed a multipath extension to AODV called Ad Hoc on-demand Multipath Distance Vector (AOMDV). AOMDV discovers multiple paths between source and destination in a single route discovery. As a result, a new route discovery is necessary only when each of the multiple paths fail. AOMDV, like AODV, ensures loop freedom and at the same time finds disjoint paths which are less likely to fail simultaneously. By exploiting already available alternate path routing information as much as possible, AOMDV computes alternate paths with minimal additional overhead over AODV.

## 3. Agent-based soft wares for simulation Ad Hoc networks

In the past, focused systems were only effective way of doing developed things. Decisions and information processes happen regular and hierarchical. But it is clear that these systems act only in the stable and unchanged company. Today, changes in technology has led companies be constantly changing and progress. Companies that can not coordinate themselves with these changes will be easily dismissed. For this reason, many companies have turned on agent-based systems. These systems use from agents that distributed on computing networks. In addition, agents not only coordinate themselves with environment, but also can learn from their environment.

However, actions and agent-based systems, are causing the evolution and don't cause impairment in systems. Though, agents, objects, Service-based architecture and etc, each are considered important technology, but their cooperation caused to create a s trong system and each alone can not have specific performance.

3.1) Agent-based tools (soft wares)

Now an increasing number of problems in industrial, commercial, medical, networking and educational application domains are being solved by agent-based solutions. These software solutions are mainly complex, open and distributed. The key abstraction in these systems is the agent. This concept is a natural metaphor similar to "object" for software engineers and they can understand, model and implement many systems on the basis of an autonomous and interacting agents. Agent-based software engineering is a powerful way of approaching large scale software engineering problems and developing agent-based systems. In this approach to software development, applications are written as agents.

Agent-based software engineering represents a novel and powerful way of approaching large scale software engineering problems. In this approach to software development, applications are written as software agents. We consider agent-based software engineering as a layered technology.

Agent-based softwares have the following features:
- complexity, openness and data control and distribution
- Usually are considered in class real-time and sensitive software.
- are formed of agents. So most important Abstraction in such systems is agent.

Various agent-based softwares are designed for network simulation, such as OPNET, NS (network simulator), GloMoSim, Swarm, NetSim and etc. In the next part, we will introduce the NS software.

3.2) Network Simulator (NS)

Software which we choose for simulation the Ad Hoc protocol is NS. Software Network Simulator (NS) is simulation software based on network events that supports the agents too. The second version of this software is named NS-2. This software provides substantial support for simulation of routing, multicast protocols and IP protocols, such as UDP, TCP, RTP and SRM over wired and wireless (local and satellite) networks.

i. **History:**
NS-2 started as a v ariant of the REAL network simulator in 1989 . REAL is a network simulator originally intended for studying the dynamic behavior of flow and congestion control schemes in packet-switched data networks.
Currently NS-2 development by VINT group is supported through Defense Advanced Research Projects Agency (DARPA) with SAMAN and through NSF with CONSER, both in collaboration with other researchers including ACIRI (see Resources). NS-2 is available on several platforms such as FreeBSD, Linux, SunOS and Solaris. NS-2 also builds and runs under Windows. Cygwin, is software that runs ns-2 on Windows.

ii. **Design**
NS-2 was built in C++ and provides a simulation interface through OTCL, an object-oriented dialect of TCL. The user describes a network topology by writing OTCL scripts, and then the main NS-2 program simulates that topology with specified parameters.
Simple scenarios should run on any reasonable machine; however, very large scenarios benefit from large amounts of memory. NS-2 requires the following packages to run: TCL, Tk, OTCL and TCLCL.

### iii. TCL

TCL (originally from "Tool Command Language", but conventionally rendered as "TCL" rather than "TCL"; pronounced as "tickle" or "tee-see-ell") is a scripting language created by John Ousterout. Originally "born out of frustration," according to the author, with programmers devising their own (poor quality) languages intended to be embedded into applications, TCL gained acceptance on its own. It is commonly used for rapid prototyping, scripted applications, GUIs and testing. TCL is used on embedded systems platforms, both in its full form and in several other small-foot printed versions. TCL is also used for CGI scripting and as the scripting language for the Egg drop bot. TCL is popularly used today in many automated test harnesses, both for software and hardware, and has a loyal following in the Network Testing and SQA communities.

The combination of TCL and the Tk GUI toolkit is referred to as TCL/Tk which is often pronounced "tickle tock."

TCL did not originally support object oriented (OO) syntax before 8.6 (8.6 provides an OO system in TCL core), so OO functionality was provided by extension packages, such as incr TCL and XOTCL. Even purely scripted OO packages exist, such as Snit and STOOOP (simple TCL-only object-oriented programming). Figure 1 simply shows NS-2 how works

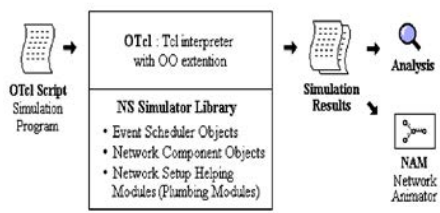

Figure 1: Simplified view of how working NS-2

### iv. Nam (Network animator)

To display the results of simulation by NS, animator software designed that called Nam. Nam is a TCL/TK based animation tool for viewing network simulation traces and real world packet trace data. The design theory behind Nam was to create an animator that is able to read large animation data sets and be extensible enough so that it could be used indifferent network visualization situations. Under this constraint Nam was designed to read simple animation event commands from a large trace file. In order to handle large animation data sets a minimum amount of information is kept in memory. Event commands are kept in the file and reread from the file whenever necessary. The first step to use Nam is to produce the trace file. The trace file contains topology information, e.g., nodes, links, as well as packet traces. Usually, the trace file is generated by NS. During an NS simulation, user can produce topology configurations, layout information, and packet traces using tracing events in NS.

### v. Nam (Network animator)

To display the results of simulation by NS, animator software designed that called Nam. Nam is a TCL/TK based animation tool for viewing network simulation traces and real world packet trace data. The design theory behind Nam was to create an animator that is able to read large animation data sets and be extensible enough so that it could be used indifferent network visualization situations. Under this constraint Nam was designed to read simple animation event commands from a large trace file. In order to handle large animation data sets a minimum amount of information is kept in memory. Event commands are kept in the file and reread from the file whenever necessary. The first step to use Nam is to produce the trace file. The trace file contains topology information, e.g., nodes, links, as well as packet traces. Usually, the trace file is generated by NS. During an NS simulation, user can produce topology configurations, layout information, and packet traces using tracing events in NS.

However any application can generate a Nam trace file. When the trace file is generated, it is ready to be animated by Nam. Upon startup, Nam will read the trace file, create topology, pop up a window, do layout if necessary, and then pause at time 0. Figure 2 shows the Nam tool environment.

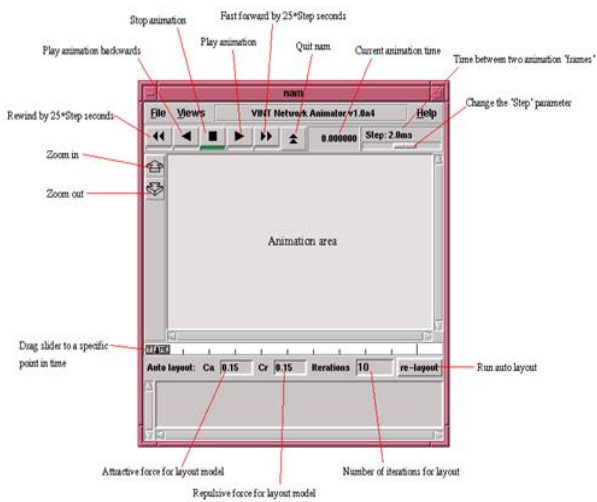

Figure 2: Nam tool environment

Xgraph also is another one of the packages in NS (and NS-2) that is used for charting the simulation results

### 3.3) Reasons for utilize the NS-2

It has many advantages that make it a useful tool, such as support for multiple protocols and the capability of graphically detailing network traffic. Additionally, NS-2 supports several algorithms in routing and queuing. LAN routing and broadcasts are part of routing algorithms. Queuing algorithms include fair queuing, deficit round-robin and FIFO.

This software also can be applied for various purposes such as the following:
- evaluate the efficiency of an existing network protocols
- Performance evaluation of new protocols designed before the implementation on the net
- perform experiments with large scale, which are not as practical possible.
- simulating the behavior of different types of IP-based networks

## 4) CASE STUDY

In this paper, we want simulate the AODV and DSDV protocols with NS-2 and compare the results. First Must write relative codes to these two protocols in a simple editor (like WordPad) and save them with TCL suffix. To write TCL codes, requires a brief acquaintance with this language which is already mentioned. After writing the relevant codes, should run them with the NS-2. For this, open Cygwin and after writing the command ns, write code name and press the enter key. With this, the TCL compiler compiles the code and NS-2 produces output files. The result of performance code can be seen visually with Nam software. For this we use the command nam. You can see view the AODV code in figure 3.

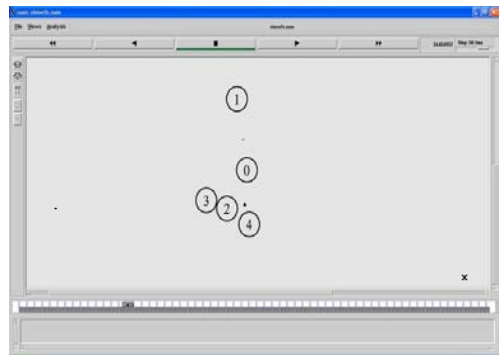

Figure 3: The result of performance the AODV code with Nam software

With pressing the play icon, simulation starts and can be seen how the protocol works.

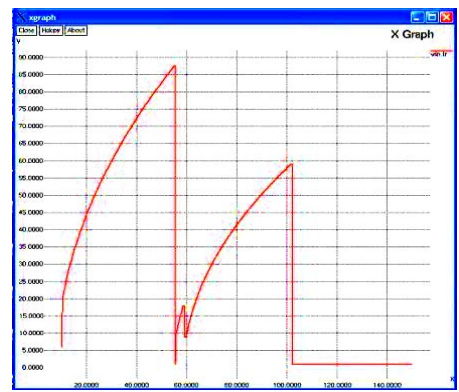

Figure 4: The AODV simulation graph

With using the command xgraph, can see the results as graphs. You can see the AODV graph in figure 4.
Implementation the DSDV code is also similarly. The relevant graph of the DSDV code shown in figure 5.

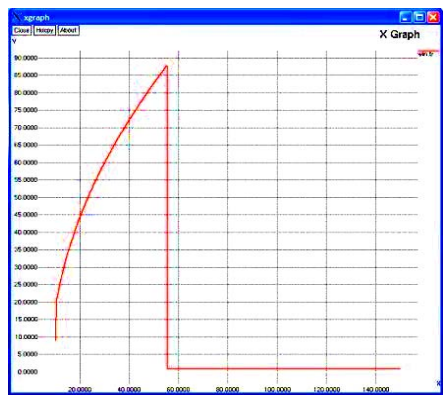

Figure 5: The DSDV simulation graph

## 5. Comparison of simulation results

As described, the AODV protocol is one of the second method protocols (On-Demand); This means that in this protocol every time a path is needed, is discovered. So, in the relevant code of this protocol that we want to communicate between nodes 4 and 1, with starting time of the simulation node 4 according to the available routing tables, detects direct path to node 1 optimum path and as you see in figure 4, at 10 starts communication.

This Process and sending data will continue until at 55 because node 1 is out of it's radio range, node 4 uses node 3 for communication (optimum path). This process continues until 59 and at this time node 4 according the routing tables and radio range of nodes, detects node 0 the optimum path to node 1 and uses it to communicate. This process also continues until 103, until node 0 gets out of radio range node 4 too and at this time as regards node 1 is also outside of radio range nodes 2 and 3, node 4 does not find no optimum path to node 1 and disconnects communication.

But, the DSDV protocol that is one of the first method protocols (Table-Driven); In this protocol, optimum paths are predetermined. This mean the optimum path between nodes 4 and 1, before starting the simulation time is defined according to the routing tables of nodes. Starting time of the simulation, we see that the direct path is diagnosed optimum path between nodes 4 and 1 and at 10, node 4 starts communication with node 1. This communication continues until 55, until at this time, because node 1 gets out of radio range node 4 communication between them is disconnected and until the end time of the simulation, another communication is not established between them; Because, as we said the optimal path between them is predetermined and during the simulation time, no attempt accomplished to find the optimal path between them.

## 6. Conclusions

Mobile Ad Hoc networks are actually the future of wireless networks; because they are cheap, simple and supple and have simple usage. We live in world that networks constantly change in it and change their topology for linkage new nodes, for this reason we go toward this networks. Notwithstanding security problems that they have, have many usages; actually their efficiency increased daily and decreased their price, for this reason they have many partials in market.

Using Ad Hoc networks for big and complex places needed high accuracy, because it is possible that this project failed and leave heavy costs. So, it is better before implement Ad Hoc for them, present a simulated model and observe the results. With this work, can see the available defects in simulated model and modify them.

There are various simulator for this work, but can say that NS-2 have special station between them; Because it is free and open source and also use it some simple. One of the it's profits is that it supports from the many protocols of wireless networks and another profit is ability mobility of nodes.

At the end, can say Ad Hoc networks have wide usage in modern world and usage of them will generalize.